\RequirePackage{fix-cm}
\documentclass[twocolumn]{svjour3}          
\smartqed  
\usepackage{graphicx}
\usepackage{newtxtext,newtxmath}
\usepackage{color}
\usepackage{subfigure} 
\usepackage{comment}   
\usepackage{natbib}
\usepackage{hyperref}
\usepackage{ulem}
\hypersetup{
    breaklinks = true,
    colorlinks = true,
    linkcolor  = blue,
    citecolor  = blue
    }

\journalname{Experiments in Fluids}
\begin{document}

\title{
Time-resolved phase-lock pressure-sensitive paint measurement of trailing edge noise dynamics
}


\author{Masato Imai$^\text{\ref{lab:a1}}$ \and
        Kohei Konishi$^\text{\ref{lab:a1}}$ \and
        Keita Ogura$^\text{\ref{lab:a1}}$ \and
        Kazuyuki Nakakita$^\text{\ref{lab:a2}}$ \and
        Masaharu Kameda$^\text{\ref{lab:a1}}$
}


\institute{Masato Imai \at
              \email{m-imai@st.go.tuat.ac.jp}           
           \and
           Masaharu Kameda \at
                \email{kame@cc.tuat.ac.jp}       
           \begin{enumerate}
                    \item Department of Mechanical Systems Engineering, Tokyo University of Agriculture and Technology, Koganei, Tokyo, Japan\label{lab:a1}
                    \item Fundamental Aeronautics Research Unit, Japan Aerospace Exploration Agency, Chofu, Tokyo, Japan\label{lab:a2}
            \end{enumerate}
}

\date{Received: date / Accepted: date}

\onecolumn
\maketitle
\section*{Abstract}
Pressure-sensitive paint (PSP) was applied to the surface of a NACA0012 airfoil to investigate pressure fluctuations associated with trailing edge (TE) noise under low-velocity flow conditions.
The primary focus is to assess the feasibility of employing laser pulses exposed at the airfoil surface to mitigate TE noise. 
However, the weak pressure fluctuations accompanying TE noise pose a challenge, as they are overshadowed by image sensor noise in high-speed cameras capturing PSP emission changes. 
To address this issue, a novel time-resolved phase-locking technique was introduced, utilizing the signal from a semiconductor pressure transducer at the trailing edge as a phase-lock trigger source. 
By repetitively conducting phase-locked measurements (1150 times), time-series ensemble-averaged data based on PSP emission images were obtained, enabling the capture of these subtle pressure fluctuations. 
Quantitatively, fluctuations with a dominant frequency of 679 Hz and an amplitude of 50 Pa are resolved within an accuracy of about 15 Pa, achieved at a recording rate of 19.2 kHz. 
Both the suppression and subsequent redevelopment of the pressure field with the TE noise offer valuable insights into the dynamics of TE noise and open avenues for targeted noise reduction strategies in aerodynamic applications.


\newpage
\null
\section*{Graphic abstract}
\begin{figure}[ht]
    \includegraphics[width=17cm]{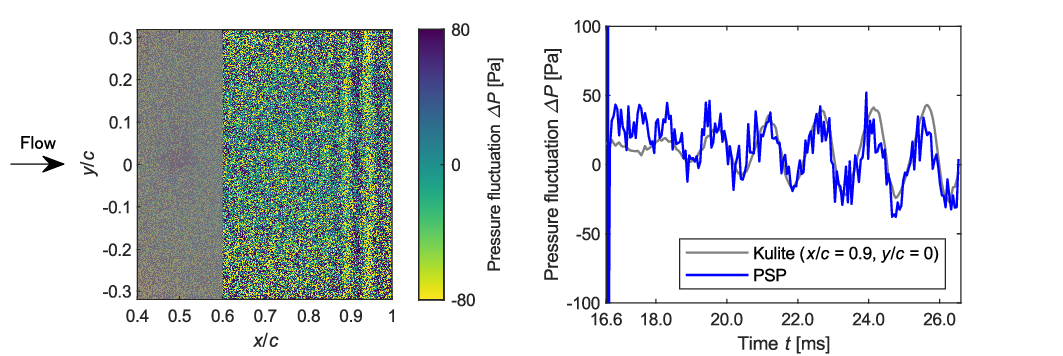}
\end{figure}

\twocolumn
\section{Introduction}
\label{sec:intro}
Airframe noise constitutes the majority of noise experienced on the ground during aircraft landings \citep{Bertsch2019}. 
Extensive efforts have been directed towards understanding the mechanisms underlying noise generation and mitigating aircraft noise using device-based technologies \citep{Molin2019, Yamamoto2019}.

Trailing-edge (TE) noise is a significant contributor to airframe noise. 
This noise, characterized by vortex sound occurring at specific angles of attack and Reynolds numbers on aircraft wings, is known to exhibit discrete tones \citep{Paterson1973}. 
Experimental and numerical studies have demonstrated the involvement of a feedback loop in the mechanism of TE noise generation \citep{Nash1999, Desquesnes2007}. 
In this loop, the detached region of the laminar boundary layer generates vortex sound upon shedding from the trailing edge, which subsequently interacts with the boundary layer. 
This interaction destabilizes the Tollmien-Schlichting (TS) wave and sustains the periodic shedding of vortices.

Various methods have been devised for controlling TS waves \citep{Simon2016, Wylie2021}, and among these strategies, controlling TS waves has also been explored as a means to reduce TE noise \citep{Inasawa2013}.
Recently, \cite{Ogura2023} proposed a non-invasive noise reduction method involving intermittent deposition of laser pulses near the boundary layer. 
Their CFD analysis and pressure measurements with semiconductor pressure transducers in wind tunnel tests suggest that this method disrupts TS waves by generating hot spots with the laser, resulting in TE noise suppression. 
A more detailed analysis of the temporal variation of TS waves is needed to elucidate the underlying mechanism.

Pressure fluctuations on the airfoil surface associated with TS waves serve as a valuable indicator to evaluate the effectiveness of control methods. 
Pressure-sensitive paint (PSP), renowned for its ability to measure pressure fields, has been extensively employed in aerodynamic research \citep{Gardner2014, Crafton2017, Elliott2022}. 
Operating as an optical pressure sensor through oxygen quenching, PSP offers high spatial resolution in capturing surface pressure distributions \citep{Liu2005}.

In this study, we utilized unsteady PSP measurements to elucidate the mechanism of the TE noise reduction method proposed by \cite{Ogura2023}. 
This attempt requires some experimental measurements.
Firstly, obtaining the unsteady pressure field over the entire airfoil surface is crucial to identifying regions where laser pulses suppress pressure fluctuations. 
Secondly, elucidating how laser pulses alter the flow field on the airfoil surface and suppress pressure fluctuations is imperative. 
Additionally, capturing the reemergence of TE noise after the cessation of the laser pulse effect is essential.

To experimentally measure the aforementioned aspects, PSP measurements capturing the time evolution of the unsteady flow field with small pressure fluctuations are indispensable. 
However, detecting pressure fluctuations caused by boundary layer instabilities such as TS waves, which have very small amplitudes on the order of 10 Pa, poses a significant challenge in PSP methodology \citep{Gregory2013, Peng2020, Goessling2020}.

Frequency domain techniques such as fast Fourier transform (FFT) and coherent output power (COP) are effective for analyzing time-series PSP image data captured by high-speed cameras \citep{Nakakita2011, Noda2018}. 
While these techniques enable visualization of pressure fluctuations by eliminating the noise in the frequency domain, they fail to capture the temporal dynamics of small pressure fluctuations induced by intermittent control methods.

Phase-lock measurement \citep{Gregory2006, Yorita2010} emerges as a suitable approach for measuring small pressure fluctuations in the time domain. 
This method reduces noise by synchronizing PSP luminescence images with pressure transducer signals, followed by ensemble averaging. 
Another method, conditional image sampling \citep{Asai2011}, determines the phase based on the timing of image capture and pressure fluctuations recorded by the pressure transducer before averaging the data.
For example, in the experiment by \citet{Gregory2006}, a pressure fluctuation vibrating at 1.3 kHz with a magnitude of 500 Pa was measured with an accuracy of 12.5 Pa.

In this research, we employed a ``time-resolved phase-lock method" to measure the temporal evolution of the flow field with small pressure fluctuations. 
The method is based on phase-locked measurements synchronized to pressure fluctuations on the airfoil surface, and laser pulses are also generated phase-locked to the same pressure fluctuations.
Hence, it allows ensemble phase averaging of a large number of pulse emission events.
The ensemble phase averaging reduces camera noise and captures small pressure fluctuations while analyzing the time evolution of the flow field in the time domain.
Given the high periodicity of TE noise pressure fluctuations at the airfoil's trailing edge \citep{Nakakita2011, Ogura2023}, the time evolution of the observed phenomenon will be the same by taking equal time intervals from a single trigger, allowing for effective ensemble averaging.

This paper is structured as follows: Section \ref{sec:experiment} describes the wind tunnel test setup and measurement methods employed. Section \ref{sec:data_pro} outlines the methods for processing measurement data. Results and discussion are presented in Section \ref{sec:results}, followed by the conclusions of this study in Section \ref{sec:conclusion}.

\section{Experiment}
\label{sec:experiment}

\begin{figure}[ht]
    \subfigure[Schematic diagram of the experimental setup in low-turbulence wind tunnel]{
        \includegraphics[width=8cm]{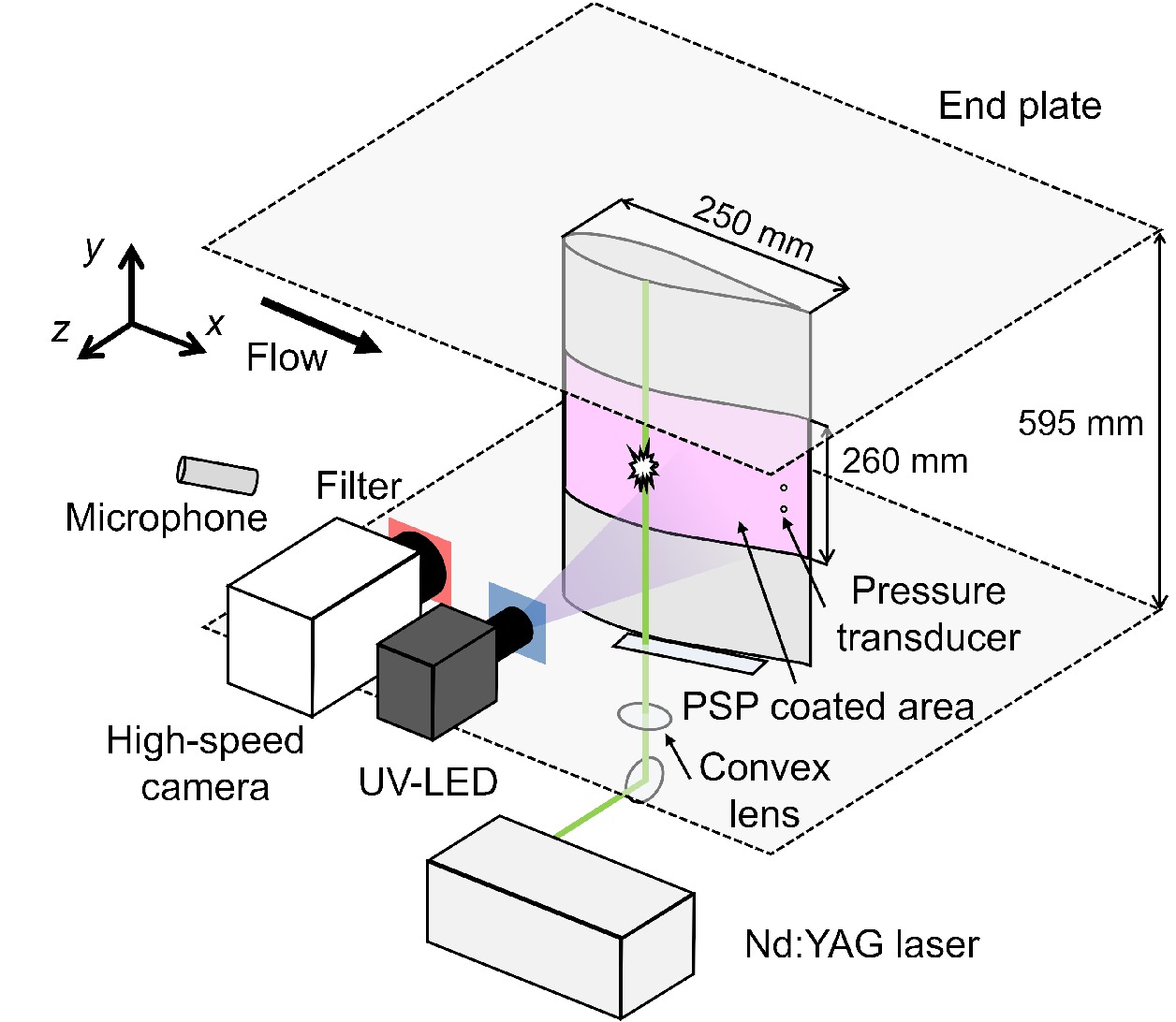}
        \label{fig:setup}
    }
    \\
    \subfigure[Cross-section view of the airfoil]{
        \includegraphics{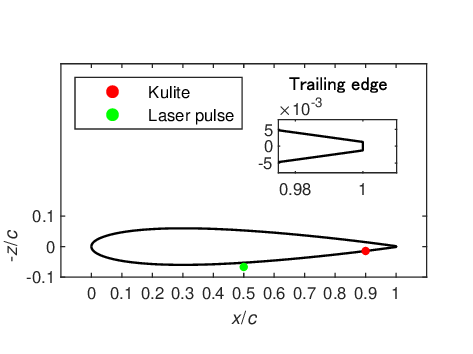}
        \label{fig:setup_airfoil}
    }
    \caption{Schematic diagrams of wind tunnel tests}
    \label{fig:setup_all}
\end{figure}

\textbf{\subsection{Wind tunnel test}
\label{subsec:wtt}}
The overall schematic of the experiment is shown in Fig. \ref{fig:setup} in which the definition of coordinates used in this paper is also indicated.
The wind tunnel tests were conducted for a two-dimensional NACA0012 airfoil model in the 0.65 m $\times 0.55$ m low-turbulence wind tunnel (LWT3) at the Chofu Aerospace Center of the Japan Aerospace Exploration Agency (JAXA).
The measurement system consists of two semiconductor pressure transducers for the acquisition of pressure reference value, a microphone for the measurement of TE noise frequency, and an optical system for the acquisition of PSP image.
A Nd:YAG laser was used for flow field control aimed at TE noise suppression. 
We defined $x$ as the direction of the free stream, $y$ as the direction along the span of the model, and $z$ as the direction perpendicular to the model surface.

The model and its placement in the wind tunnel test section followed the experiments of \cite{Ogura2023}.
The model was mounted vertically on an open test section, with end plates at both ends of the span.
The model was made of an aluminum alloy (Japanese Industrial Standards (JIS) A5052), with dimensions of chord length $c = 250$ mm and span length $b = 595$ mm.
The cross-section of the wing model is shown in Fig. \ref{fig:setup_airfoil}.
Its trailing edge has a blunt profile and is 0.63 mm thick.
For the PSP measurements, the pressure side of the wing model was coated with polymer-ceramic PSP (PC-PSP) \citep{Sugioka2018}, which consists of platinum tetra (pentafluorophenyl) porphine (PtTFPP) as the pressure-sensitive dye, an ester-based polymer, and titanium dioxide particles as the ceramic component. 
The response frequency of this PC-PSP is 3 kHz \citep{Sugioka2018}.

The two unsteady semiconductor pressure transducers (XCQ-093-5D, Kulite) were embedded in the pressure side at $x/c = 0.9$, both at spanwise positions $y/c = 0$ and $y/c = -0.2$, as shown in Fig. \ref{fig:setup_airfoil}.
The microphone (Type 4958, Brüel \& Kjær) was placed on the pressure side of the airfoil.
The data acquisition system and the phase-locked system with the signal from the pressure transducer are explained in Section \ref{subsec:measure}.

For PSP measurements, excitation light from an ultraviolet LED (UV-LED; IL-106X, Hardsoft, emission wavelength: 390 nm) was applied to the code region from $x/c = 0.3$ to $1.0$ in the model.
The luminescence of PSP was captured by a 12-bit CMOS high-speed camera (Phantom v1840, Vision Research). 
A cold filter (SC0451, Asahi Spectra, cut-off wavelength: 450 nm) was attached to the UV-LED, and the high-speed camera was equipped with a lens (Nikkor 50 mm f/1.2, Nikon) along with a sharp cut long-pass filter (SCF-50S-58O, Sigma Koki, cut-on wavelength: 580 nm) to capture the PSP luminescence.

The Nd:YAG laser (SAGA 230, Thales) operates at an output wavelength of 532 nm and a repetition frequency of 10 Hz, with a pulse width ranging from 4 to 8 ns.
Positioned outside the test section, its beam was focused approximately 3.5 mm above the wing surface at $x/c = 0.5$, $y/c = 0$ on the pressure side, as shown in Fig. \ref{fig:setup_airfoil}, using mirrors and a convex lens (SLB-30-500PM, Sigma Koki, focal length: 500 mm), generating laser-induced plasma.
The focused point of the laser was outside of the boundary layer (about 1.5 mm thick), and its chord position was suitable for TS wave disruption \citep{Ogura2023}.

\textbf{\subsection{Time-resolved phase-lock measurement}
\label{subsec:measure}}
The PSP measurements were conducted in synchronization with the phase of pressure fluctuations captured by the reference pressure transducer. 
To achieve this synchronization, it was necessary to generate a trigger signal aligned with a specific phase of the transducer signal. 
Subsequently, the high-speed camera was operated using this trigger signal. 
Given the presence of electrical noise in the reference pressure transducer signal, signal filtering was essential for accurately identifying the phase of the fluctuations and creating an appropriate phase-locked measurement signal \citep{Gregory2006,Yorita2010}.

\begin{figure}[ht]
    \subfigure[Signal generating system]{
        \includegraphics[width=6.5cm]{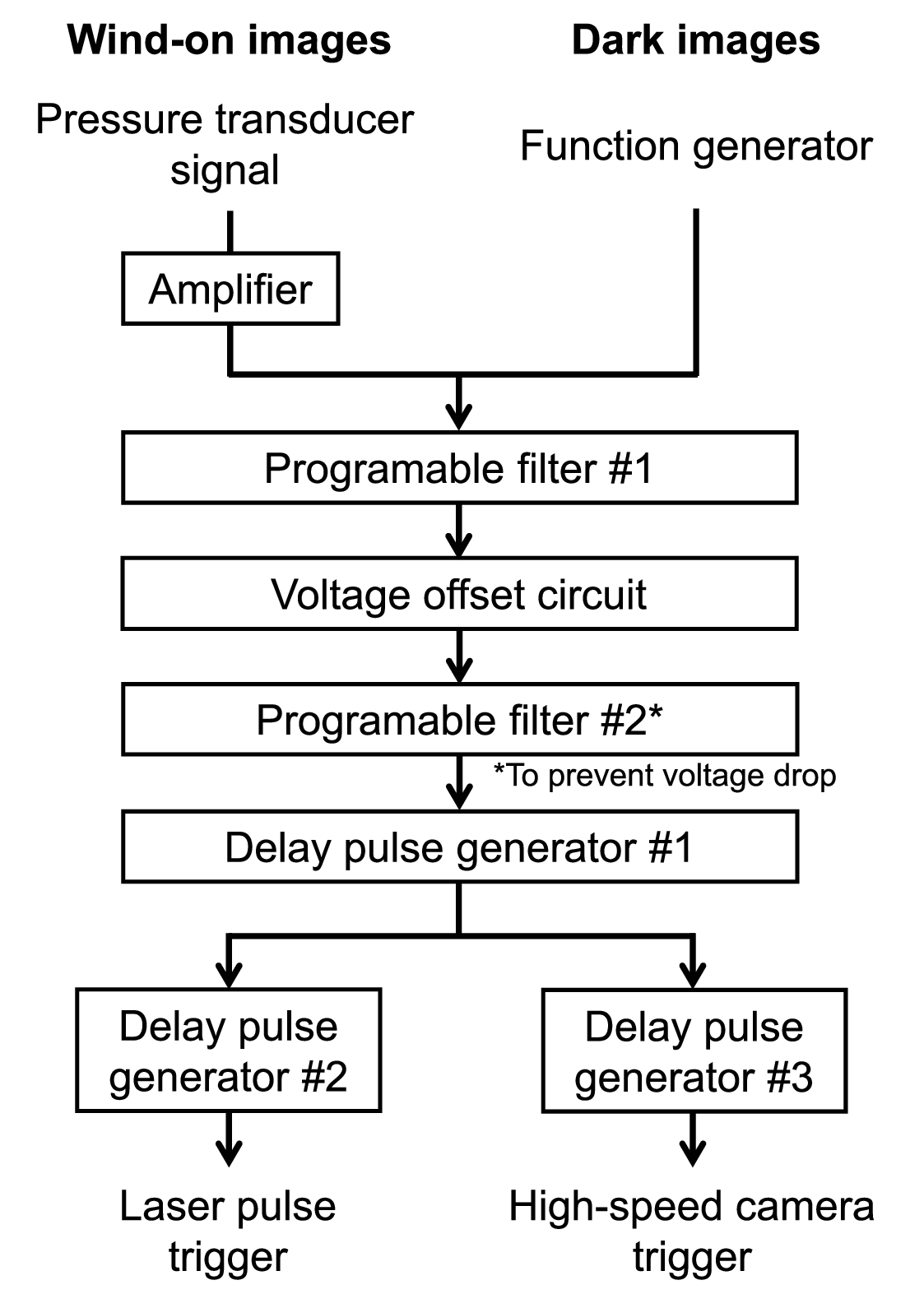}
        \label{fig:filter}
    }
    \\
    \subfigure[Schematic image of the signal for phase-lock measurement in PSP]{
        \includegraphics[width=8cm]{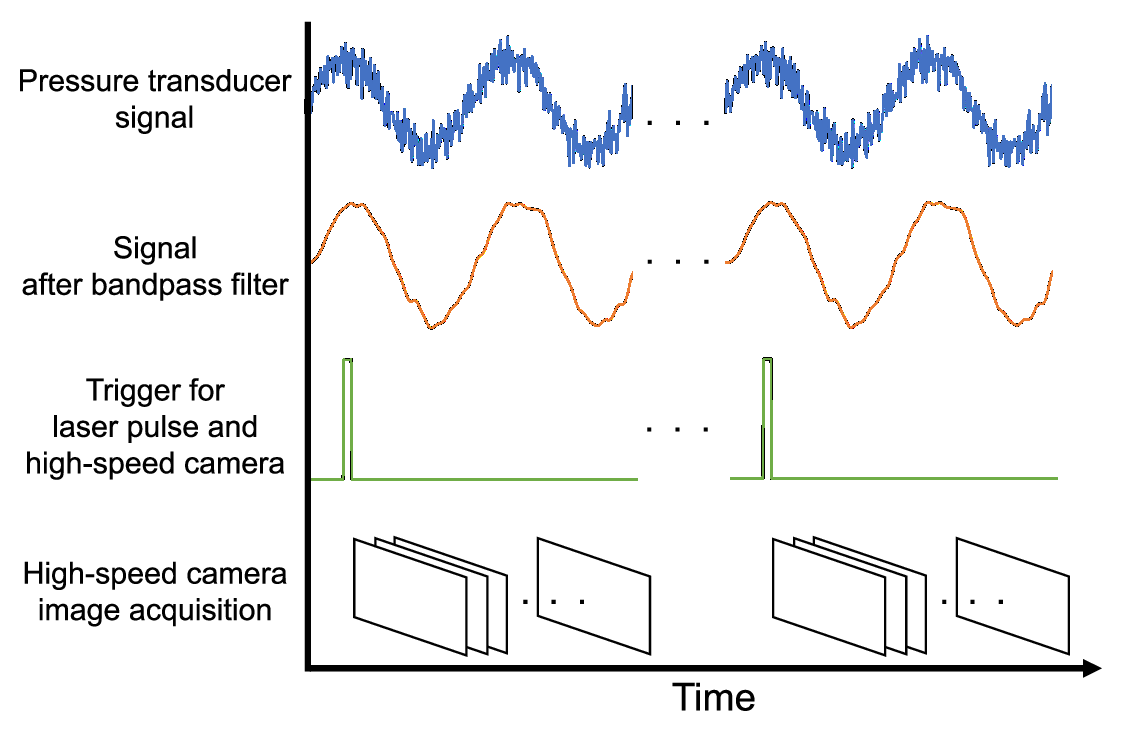}
        \label{fig:signal}
    }
    \caption{Signal system of time-resolved phase-lock measurement}
    \label{fig:filter_signal}
\end{figure}

The signal system utilized for the time-resolved phase-lock PSP measurement in this study is illustrated in Fig. \ref{fig:filter_signal}. 
The signal source was the semiconductor pressure transducer signal at the trailing edge of the wing ($x/c = 0.9$, $y/c = 0$), which was through a DC amplifier (AM30, Unipulse). 
A programmable filter (3628, NF Corp.) was employed as a bandpass filter to eliminate noise from the source signal. 
The bandpass filter was configured with a high-pass filter set at 0.40 kHz and a low-pass filter set at 1.50 kHz, specifically selected to isolate the frequency range associated with TE noise. 
Subsequently, the signal underwent conditioning through an in-house voltage offset circuit and another programmable filter (3627, NF Corp). 
These devices were utilized to fine-tune the voltage to the triggering threshold and prevent voltage drop. 
The filtered signal was then employed as the reference for phase-locking, while a delay pulse generator (9520, Quantum Composers) generated fundamental triggers for both the laser and the high-speed camera. 
Due to the laser's maximum repetition rate of 10 Hz, the trigger counter function of the delay pulse generator was utilized to produce one trigger signal after every 70 external trigger signals were received. 
This approach effectively reduced the occurrence frequency of trigger signals aligned with the frequency of TE noise to a repetition rate of 10 Hz for the laser.
The second delay pulse generator (DG535, Stanford Research Systems) was used to generate two triggers for operating the Nd:YAG laser.

Additionally, to capture the inherent noise of the high-speed camera's image sensor, dark images were obtained by substituting the semiconductor pressure transducer signal with one from a function generator (WF1974, NF Corp.), set to the same frequency as the TE noise, as illustrated in Fig. \ref{fig:filter}.

In this study, we used the intensity-based method for PSP measurements. 
The high-speed camera was configured to capture the PSP's luminescence continuously in response to a single trigger signal, as shown in Fig. \ref{fig:signal}.
It captured a series of 192 images at a frame rate of 19.2 kHz, with each exposure time lasting 49.79 $\mu$s. 
This configuration allowed for approximately 10 ms of measurement per trigger. 
The imaging sequence was conducted over 1150 triggers, resulting in a total of 220,800 images, with the entire measurement process spanning approximately 140 s.
The high-speed camera operated in binning mode to reduce noise, capturing images at a resolution of $768 \times 576$ pixels.
This setup achieved a spatial resolution of 0.28 mm per pixel for the PSP images. 

In \citet{Ogura2023}'s experiments, time-series data from a semiconductor pressure transducer at the airfoil's trailing edge showed that TE noise pressure fluctuations are suppressed around 6 ms after laser pulse application, with the suppression lasting for a period before pressure fluctuations start to redevelop approximately 16 ms post-application.
This experiment focused on PSP measurements at two critical phases:
\begin{enumerate}
    \renewcommand{\labelenumi}{(\roman{enumi})}
    \item Suppression of pressure fluctuations
    \item Redevelopment of pressure fluctuations
\end{enumerate}
To synchronize the capture timing with these phases, we used the third delay pulse generator (WF1944, NF Corp.) to adjust the generation of the capture trigger signal, as depicted in Fig. \ref{fig:filter}.

The semiconductor pressure transducer and microphone signals were recorded at a 50 kHz sampling rate using a digital oscilloscope (DL850, Yokogawa). 
To compare with the PSP data, the time-series data from the semiconductor pressure transducers were downsampled to match the high-speed camera's sampling rate after being ensemble-averaged 256 times.

\section{Image data processing}
\label{sec:data_pro}
The data processing procedure for PSP images is outlined in detail. 
Given the focus on measuring extremely weak pressure fluctuations, considerable effort was invested in minimizing noise originating from the camera. 
The random component of the noise was reduced through ensemble averaging of the time-series data and spatial binning. 
In parallel, the bias component of the noise was addressed by subtracting timing-dependent dark noise and conducting \textit{in situ} calibration using the pressure transducer on the airfoil.

\begin{figure}[ht]
    \includegraphics[width=5cm]{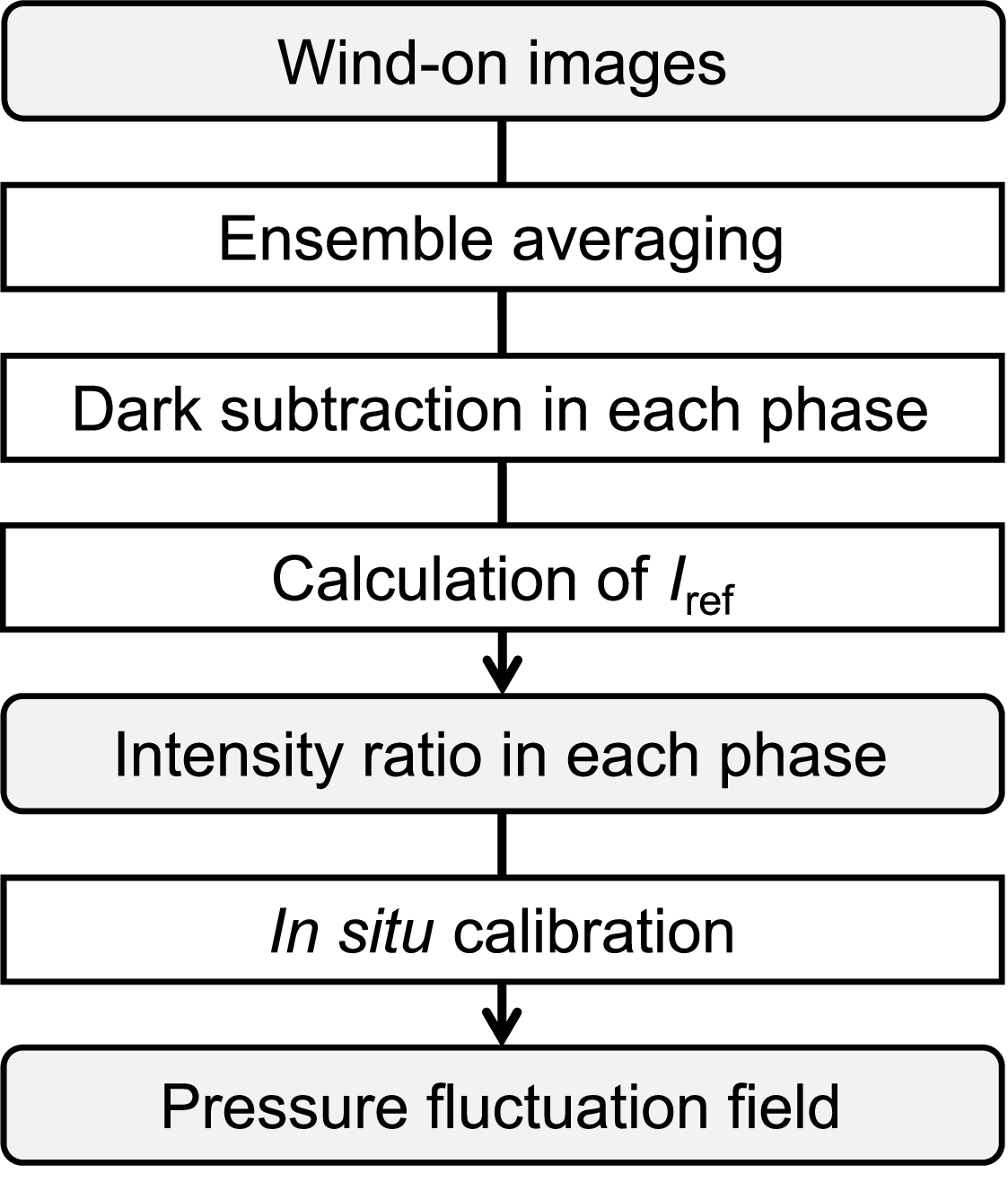}
    \caption{Data processing flow for phase-lock PSP measurement data}
    \label{fig:data_flow}
\end{figure}

The data processing flow to calculate the temporal evolution of small pressure fluctuations from PSP image data is depicted in Fig. \ref{fig:data_flow}. 
All data processing steps were performed using MATLAB. 
First, 1150 sets of wind-on (run) PSP images consisting of 192 consecutive frames were ensemble averaged at each shooting frame timing (phase). 
Similarly, the same process was performed for 300 sets of dark images to capture the dark noise of the image sensor. 
The ensemble-averaged PSP images obtained at each frame timing were then subjected to dark subtraction using the ensemble-averaged dark images at the corresponding frame timing to remove biased image sensor noise and obtain only the PSP emission intensity $I$. 

\begin{figure}[ht]
    \includegraphics{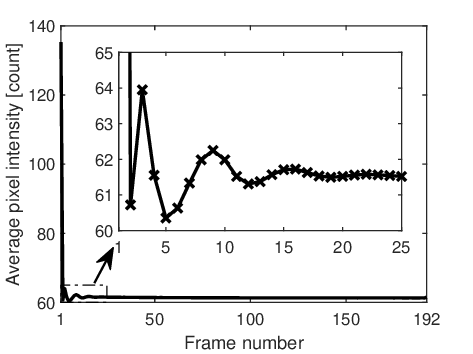}
    \caption{Average intensity of all pixels at each capture timing of dark images (The figure in the upper right corner of the graph is an enlargement of the dashed section within the graph.)}
    \label{fig:dark_plot}
\end{figure}

It was important in this study to perform dark subtraction of the wind-on image at each frame timing.
Figure \ref{fig:dark_plot} shows the average luminescence intensity within the dark image at each frame timing, where the dark values fluctuate until the 20th frame and then stabilize. 
Although the fluctuations in dark values are subtle compared to the 12-bit (4096) resolution of the high-speed camera, they significantly affect the pressure calculation for the weak pressure fluctuations targeted in this PSP measurement. 

For calculating pressure using the intensity method, the reference image $I_{\rm{ref}}$ was defined as the average of wind-on images, excluding frames influenced by the laser pulse and the significantly noisy last frame (frame 192).
The average was determined from images spanning frames 4 to 191.
Based on this average, the intensity ratio $I_{\rm{ref}}/I$ was calculated for each frame timing.

Using the time-series of $I_{\rm{ref}}/I$, we extracted and averaged values from 44 pixels around the pressure transducer at $x/c = 0.9, y/c = -0.2$.
This average, combined with time-series data from the pressure transducer, allowed us to derive the PSP pressure calibration coefficients using the \textit{in situ} method. 
Frames 10 to 191 were selected for calculating these coefficients, due to reduced variation in dark levels within this range, as evidenced in Fig. \ref{fig:dark_plot}. 
The selection of this pressure transducer position is because it was not influenced by the laser pulse and consistently exhibits pressure fluctuations at the trailing edge of the wing.
The PSP calibration coefficients were determined using the empirical equation shown in Eq. (\ref{eq:dP}) as
\begin{equation}
    \it{\Delta} P = A + B\frac{I_{\rm{ref}}}{I}.
    \label{eq:dP}
\end{equation}
Here, $\it{\Delta} P$ represents pressure fluctuation values from the pressure transducer, and $A$ and $B$ are the calibration coefficients for the PSP.
This formulation assumes a uniform temperature distribution on the wing surface and a consistent pressure sensitivity of the PSP.
By applying these PSP calibration coefficients in data processing, the field of temporal pressure fluctuations was calculated.

The residual noise in the images, after performing dark subtraction, primarily stems from the camera's shot noise.
According to \citet{Liu2001}, this type of noise can be reduced by the square root of the number of images averaged, denoted as $N^{1/2}$. 
This reduction method is equally effective for spatial averaging.
In this study's data processing, we first conducted spatial binning over $2\times2$ pixels during image capture and then averaged across 1150 repetitions. 
Consequently, the repetition count, $N_{\rm rep} = 2\times2\times1150$, leads to a noise reduction by a factor of approximately 68 times, as calculated by ${N_{\rm rep}}^{1/2}$.
The images presented in the results section have undergone noise reduction to this extent.
Additionally, for comparison with time-series pressure values from the pressure transducer, further averaging over 44 pixels surrounding the pressure transducer was conducted, achieving a noise reduction factor of about 450 times, denoted as ${N_{\rm rep}}^{1/2}$. 

\section{Results and discussion}
\label{sec:results}

\textbf{\subsection{Experimental conditions}}
\label{sec:conditon}
The airflow conditions of the wind tunnel were set at a freestream velocity of $U_{0} = 22.7$ m/s.
The angle of attack for the wing model was set to $\alpha = 2^\circ$, under which conditions TE noise was generated. 
The Reynolds number, based on the wing chord length, was $Re = 3.8\times10^5$.
Furthermore, the Nd:YAG laser, utilized for noise suppression, operated with an average power of 0.98 W.

\textbf{\subsection{Overview of the experimental data}
\label{subsec:overview}}
This section first discusses the TE noise generated during wind tunnel tests and the changes in pressure fluctuations when laser pulses are introduced, based on results from both a microphone and a semiconductor pressure transducer.

\begin{figure}[ht]
    \includegraphics{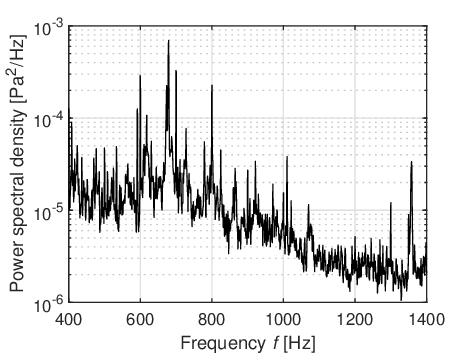}
    \caption{Power spectral density of time-series data measured by the microphone}
    \label{fig:pow_spec}
\end{figure}

The frequency of the TE noise generated from the wing model during the wind tunnel experiments was verified via the microphone's power spectrum analysis. 
The results of the power spectral density, shown in Fig. \ref{fig:pow_spec}, reveal a prominent peak at a frequency of $f = 679$ Hz, pinpointing the principal frequency of TE noise for this study.

\begin{figure}[ht]
    \includegraphics{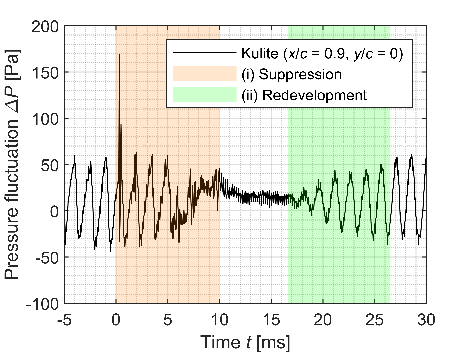}
    \caption{Time-series pressure fluctuation values at the trailing edge of the wing's pressure side ($x/c = 0.9, y/c = 0$) and the timing of high-speed camera image acquisition}
    \label{fig:timing}
\end{figure}

Figure \ref{fig:timing} shows the changes in pressure fluctuations at the pressure side of the wing, $x/c = 0.9, y/c = 0$, as measured by the semiconductor pressure sensor upon the application of laser pulses.
The time axis is set such that $t = 0$ ms corresponds to the moment when the laser pulse was applied at the wing surface location $x/c = 0.5, y/c = 0$.
The pressure fluctuations with an amplitude of about 50 Pa observed before $t = 0$ ms correspond to the pressure fluctuations at the trailing edge due to TE noise. 
These pressure fluctuations continue for a while after the application of the laser pulse but begin to be suppressed starting from $t = 5.5$ ms.
After being suppressed until $t = 18$ ms, the pressure fluctuations start to amplify, and by $t = 29$ ms, they have returned to their original amplitude. 
These changes in pressure fluctuations are consistent with the results found by \citet{Ogura2023}.

In this PSP measurement, the characteristic temporal changes in small pressure fluctuations were captured during two timings highlighted in Fig. \ref{fig:timing}:
\begin{enumerate}
    \renewcommand{\labelenumi}{(\roman{enumi})}
    \item Suppression of pressure fluctuations\\
            ($t = 0$ to $10$ ms)
    \item Redevelopment of pressure fluctuations\\
            ($t = 16.6$ to $26.6$ ms)
\end{enumerate}
Note that these measurements were conducted in separate wind tunnel runs.

Given the TE noise frequency of $f = 679$ Hz, which corresponds to a period of $T = 1.47$ ms, the frame rate of our high-speed camera enables capturing one cycle of TE noise pressure fluctuations in approximately 28 phases (about $13^{\circ}$ each), effectively recording around 7 cycles across 192 continuous frames.

\textbf{\subsection{Measurement accuracy}
\label{subsec:accuracy}}
The discussion here centers on the accuracy of time-resolved phase-lock measurements using PSP, specifically targeting data from the suppression process ($t = 0$ to $10$ ms).

\begin{figure}[ht]
    \includegraphics{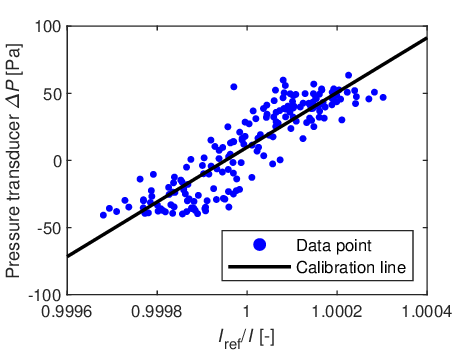}
    \caption{PSP \textit{in situ} calibration results at $x/c = 0.9, y/c = -0.2$ (During suppression process measurement. Note that at this measurement point, there is no suppression occurring in the pressure fluctuations.)}
    \label{fig:psp_1_cal}
\end{figure}

An initial evaluation of the PSP calibration coefficients obtained through the \textit{in situ} method is provided. 
Figure \ref{fig:psp_1_cal} illustrates the relationship between the pressure fluctuation values at the wing's trailing edge on the pressure side ($x/c = 0.9$, $y/c = -0.2$) as measured by the pressure transducer, and the $I_{\rm{ref}}/I$ from the PSP surrounding this area. 
The displayed PSP data are the result of ensemble averaging at each timing and spatial averaging across 44 pixels, following the procedures outlined in Section \ref{sec:data_pro}.
The observed data confirm that the time-resolved phase-lock measurement method successfully captured small pressure fluctuations within the range of -50 to 50 Pa.
By applying a linear approximation to this data, the calibration coefficients $A$ and $B$ for pressure were determined according to Eq. (\ref{eq:dP}). 

In this measurement, the luminance count of $I_{\rm{ref}}$ was around 2000, which is about half of the full count (4096). 
Based on Fig. \ref{fig:psp_1_cal}, the $I_{\rm{ref}}/I$ variation for a 50 Pa fluctuation approximates $0.025\%$, leading to an actual luminance change of roughly 0.5 count.
This highlights the importance of accounting for frame-specific luminance variances in dark images, as detailed in Section \ref{sec:data_pro}, especially when dealing with such small luminance changes.

To explore how the accuracy of pressure calculations in time-resolved phase-lock PSP measurements is influenced by the number of ensemble averages, we adjusted the ensemble averaging from 1 to 1150. 
The data processing method was consistent with that outlined in Section \ref{sec:data_pro}, except for the variation in the number of ensemble averages. 
The accuracy was assessed using the coefficient of determination $R^2$ for the PSP pressure calibration line, with time-series ensemble averaged pressure fluctuation values from the pressure transducer as the reference. 
The $R^2$ value, which ranges from 0 to 1, serves as an indicator of accuracy, with values nearer to 1 denoting greater precision in PSP pressure calculations.

\begin{figure}[ht]
    \includegraphics{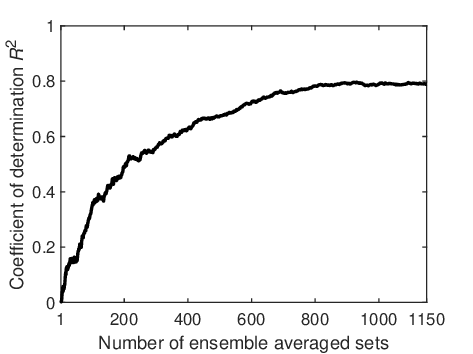}
    \caption{The effect of the number of ensemble averages on the coefficient of determination}
    \label{fig:psp_1_ensemble}
\end{figure}

The results showed an increase in $R^2$ as the number of ensemble averages increased, as shown in Fig. \ref{fig:psp_1_ensemble}. 
This indicates the effective reduction of random noise, particularly shot noise, through ensemble averaging.
In particular, it was observed that $R^2$ approaches approximately 0.8 with about 900 ensemble averages, suggesting that this level of averaging is necessary to measure the phenomenon in focus. 
The reason the coefficient of determination $R^2$ does not approach 1 suggests the presence of measurement errors beyond the shot noise from the high-speed camera. Possible sources of error are discussed subsequently.

\begin{figure*}[ht]
    \subfigure[$t = 2$ ms]{
        \includegraphics[height=5.6cm]{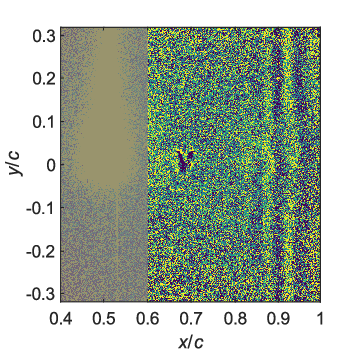}
        \label{fig:psp_1_p_1}
    }
    \subfigure[$t = 7$ ms]{
        \includegraphics[height=5.6cm]{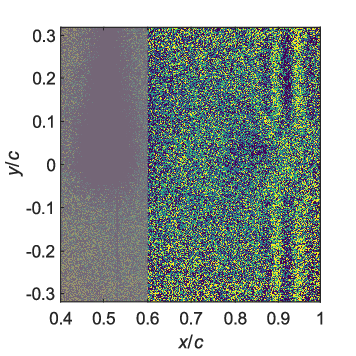}
        \label{fig:psp_1_p_2}
    }
    \subfigure[$t = 9$ ms]{
        \includegraphics[height=5.6cm]{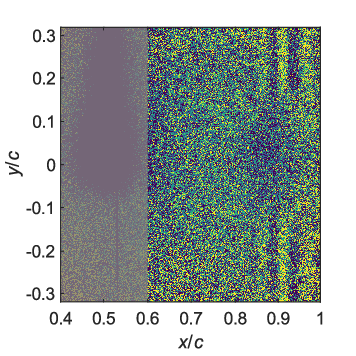}
        \label{fig:psp_1_p_3}
    }\\
    \centering
    \includegraphics[width=15cm]{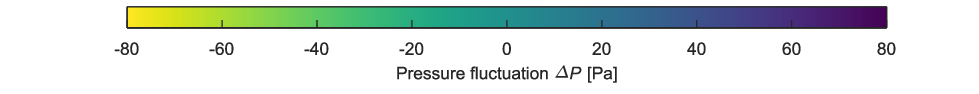}
    \caption{Visualization of surface pressure fluctuation field during suppression process (Note that the range $x/c = 0.4$ to $0.6$ represents an area, masked in gray, where pressure measurements were not obtainable due to the effects of laser pulses.)}
    \label{fig:psp_1_p}
\end{figure*}

\begin{figure*}
    \subfigure[Position at $x/c = 0.9, y/c = -0.2$]{
        \includegraphics{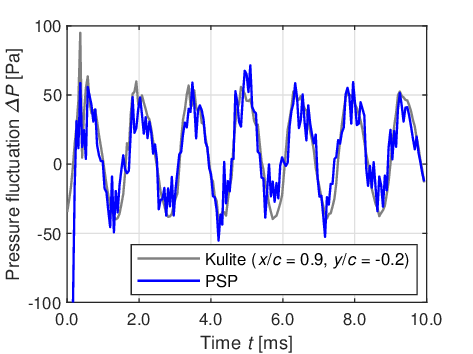}
        \label{fig:psp_1_p_k_1}
    }
    \subfigure[Position at $x/c = 0.9, y/c = 0$]{
        \includegraphics{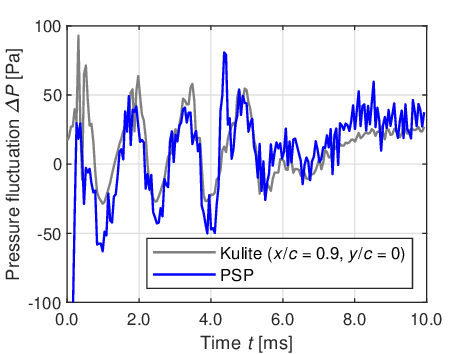}
        \label{fig:psp_1_p_k_2}
    }
    \caption{Comparison of the time-series pressure fluctuation values of PSP and pressure transducers during suppression process}
    \label{fig:psp_1_p_k}
\end{figure*}

Based on the obtained PSP pressure calibration coefficients, the pressure fluctuation fields for each frame timing were calculated. 
Representative pressure fields on the wing pressure side after the application of laser pulses at $t = 2, 7, 9$ ms are shown in Fig. \ref{fig:psp_1_p}. 
The airflow direction is positive along the $x$-axis.
The range $x/c = 0.4$ to $0.6$ and $y/c = -0.3$ to $0.3$ represents an area, masked in gray, where PSP was excited and degraded due to the effect of laser pulses. 
Consequently, the regions exhibiting extremely low (Fig. \ref{fig:psp_1_p_1}) and high (Figs. \ref{fig:psp_1_p_2} and \ref{fig:psp_1_p_3}) pressures are not indicative of the actual pressure field caused by laser pulses, but rather areas where pressure measurements were not accurately obtained.
From Fig. \ref{fig:psp_1_p_1}, a pressure fluctuation structure of about 50 Pa can be observed at the trailing edge ($x/c = 0.85$ to $0.95$) along the spanwise direction ($y$-axis).
As time progresses, as shown in Figs. \ref{fig:psp_1_p_2} and \ref{fig:psp_1_p_3}, this pressure fluctuation structure disappears within the spanwise range of $y/c = -0.1$ to $0.1$, while it is maintained at other span positions.
The disappearance of the pressure fluctuation begins around $t = 7$ ms (Fig. \ref{fig:psp_1_p_2}), starting to fade in an arc-like structure, with this arc expanding over time, increasing the area of suppression.
These results not only visualize the small pressure fluctuations at the wing's trailing edge but also successfully capture the spatial and temporal extent of the laser pulse's impact through time-resolved phase-lock measurement.

Pressure fluctuation values measured by PSP were compared with those from two pressure transducers installed at the trailing edge. 
The comparison for the position $x/c = 0.9$, $y/c = -0.2$ is shown in Fig. \ref{fig:psp_1_p_k_1}, and for $x/c = 0.9$, $y/c = 0$ is shown in Fig. \ref{fig:psp_1_p_k_2}.
It is noteworthy that the PSP pressure calibration coefficients used for this analysis were calculated via the \textit{in situ} method, based on the pressure transducer data shown in Fig. \ref{fig:psp_1_p_k_1}.

From Fig. \ref{fig:psp_1_p_k}, small pressure fluctuations at two locations are successfully captured over time using PSP.
The error over time between the PSP and the pressure transducer in Fig. \ref{fig:psp_1_p_k_1} was 14.8 Pa in terms of the root mean square (RMS) value.
For the calculation of this RMS value, frames 10 to 191 were used, as same as the PSP calibration.
The RMS value of the 50 Pa pressure fluctuation was 34 Pa, and the corresponding fluctuation in $I_{\rm{ref}}/I$ was 0.025\%, as previously discussed in Fig. \ref{fig:psp_1_cal}.
Furthermore, considering that the stability of the excitation light source was approximately 0.01\% for this measurement sequence (refer to the \hyperref[sec:appendix]{Appendix}), the expected pressure calculation error due to variations in the excitation light source was calculated as $34\times0.01/0.025 = 14$ Pa.
This value was close to the error observed between the pressure transducer and the PSP. 
Therefore, it is likely that the measurement error in this study originated from the stability of the excitation light source.

The missing values around $t = 0$ ms are due to the application of the laser pulse at that moment, which excited the PSP, making it unable to measure pressure accurately.
In Fig. \ref{fig:psp_1_p_k_1}, a state is captured where pressure fluctuations are maintained without being affected by the laser pulse.
Figure \ref{fig:psp_1_p_k_2} shows that it is possible to follow the process in which these fluctuations are suppressed after the application of the laser pulse, even with PSP measurements.
The peak observed around $t = 4.4$ ms corresponds to the high-pressure fluctuation identified at the location $x/c = 0.7$, $y/c = 0$ in the pressure fluctuation field shown in Fig. \ref{fig:psp_1_p_1}.
Given that this peak is not observed in the pressure transducer signals, it is thought not to represent actual pressure fluctuations on the wing surface. 
Instead, it is suspected to arise from the passage of a low-density area created by the laser pulse \citep{Bright2018}. 
Further discussion of this topic will be provided in subsequent sections.

\begin{figure}
    \includegraphics{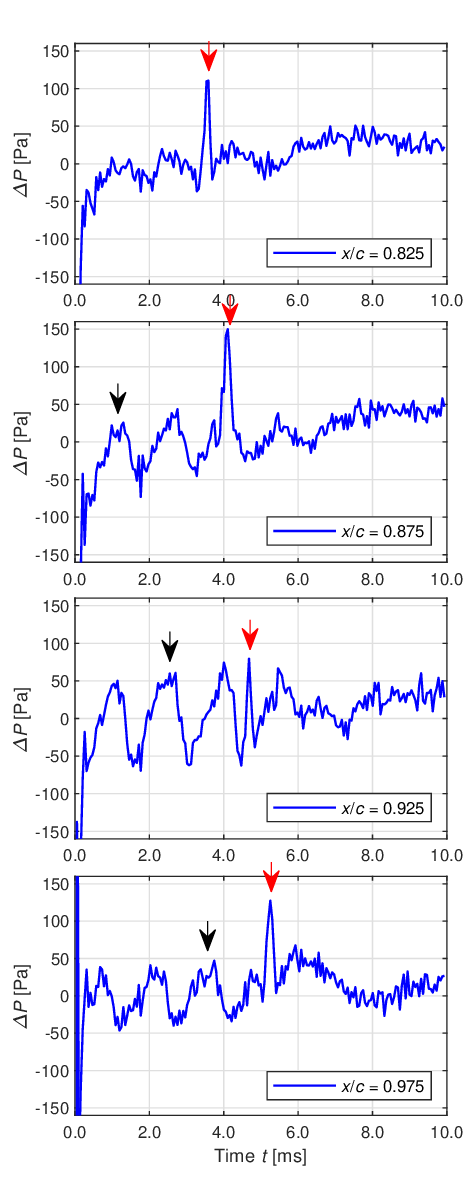}
    \caption{Time-series pressure fluctuations at various chord positions on the wing trailing edge ($x/c = 0.825$ to $0.975, y/c = 0$) during the suppression process}
    \label{fig:psp_1_te}
\end{figure}

\textbf{\subsection{Suppression process}
\label{subsec:suppress}}
Utilizing the outcomes visualized through time-series PSP measurements, a discussion on the structure of pressure fluctuations is presented.
To closely examine the time evolution of pressure fluctuations at different points on the trailing edge, data from the range $x/c = 0.825$ to $0.975$, at $y/c = 0$, with increments of 0.05, are presented in Fig. \ref{fig:psp_1_te}.

The measured phase velocity of the pressure fluctuation (marked by black arrows in Fig. \ref{fig:psp_1_te}), derived from time-series pressure field data, was $v_{\rm ph} = 8.5$ m/s. 
By applying the phase velocity calculation method introduced by \citet{Nakakita2013}—with TE noise frequency $f = 679$ Hz and a single wavelength of pressure fluctuation $\lambda = 12.6$ mm—the phase velocity is determined to be $v_{\rm ph} = f\lambda = 8.6$ m/s. 
The agreement of these values confirms the accuracy of the PSP measurement in determining velocities at the wing's trailing edge.
Such continuous, time-resolved phase-lock measurements enable a detailed evaluation of complex flow fields.

From $t = 0$ to $3$ ms, before the influence of laser pulses, pressure fluctuations at various chord locations display varying magnitudes. 
The amplitude increases moving upstream towards the trailing edge and then decreases as it nears the edge. 
After $t = 4$ ms, the effect of the laser pulses becomes apparent, suppressing pressure fluctuations progressively from the upstream side toward the trailing edge.

\begin{figure*}[ht]
    \subfigure[$t = 18$ ms]{
        \includegraphics[height=5.6cm]{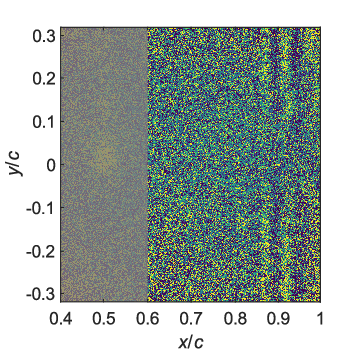}
        \label{fig:psp_2_p_1}
    }
    \subfigure[$t = 20$ ms]{
        \includegraphics[height=5.6cm]{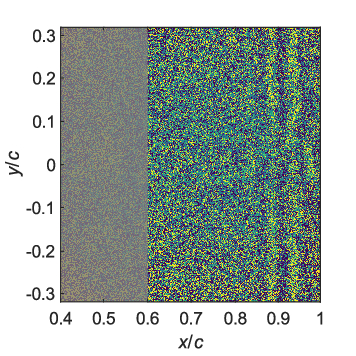}
        \label{fig:psp_2_p_2}
    }
    \subfigure[$t = 26$ ms]{
        \includegraphics[height=5.6cm]{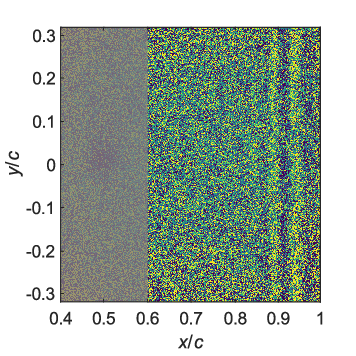}
        \label{fig:psp_2_p_3}
    }\\
    \centering
    \includegraphics[width=15cm]{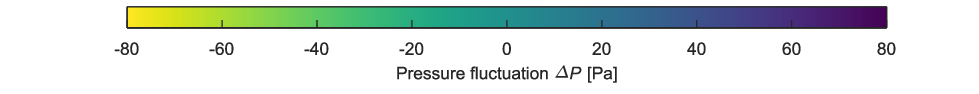}
    \caption{Visualization of surface pressure fluctuation field during redevelopment process (Note that the range $x/c = 0.4$ to $0.6$ represents an area, masked in gray, where pressure measurements could not be obtained accurately due to the effects of laser pulses.)}
    \label{fig:psp_2_p}
\end{figure*}

\begin{figure*}[ht]
    \subfigure[Position at $x/c = 0.9, y/c = -0.2$]{
        \includegraphics{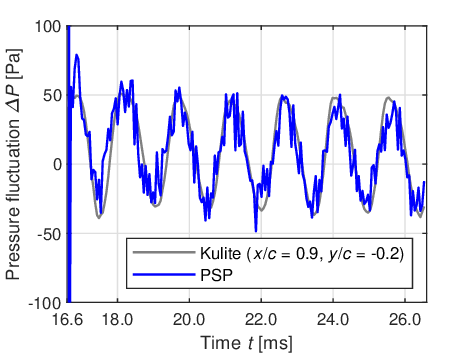}
        \label{fig:psp_2_p_k_1}
    }
    \subfigure[Position at $x/c = 0.9, y/c = 0$]{
        \includegraphics{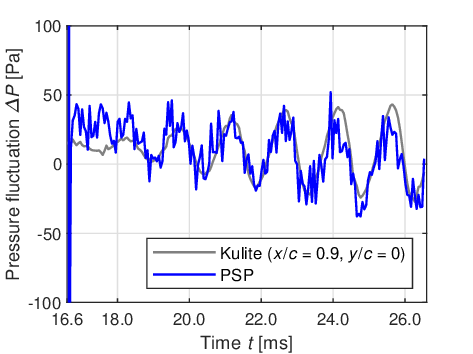}
        \label{fig:psp_2_p_k_2}
    }
    \caption{Comparison of the time-series pressure fluctuation values of PSP and pressure transducers during redevelopment process}
    \label{fig:psp_2_p_k}
\end{figure*}

The peaks observed at each chord location in Fig. \ref{fig:psp_1_te} (marked by red arrows) align with similar peaks seen in Fig. \ref{fig:psp_1_p_k_2} and are also visualized in Fig. \ref{fig:psp_1_p_1}.
As briefly discussed earlier, the absence of these peaks in the pressure transducer data implies that they may not represent actual surface pressure fluctuations but are instead pressure inaccuracies due to the PSP measurement.
In this study, the PSP pressure calculation employs the $I_{\rm ref}/I$; thus, an increase in pressure values corresponds to a reduction in luminance values in the actual measurement image $I$ at these timings.
Since PSP is sensitive not only to pressure but also to temperature \citep{Liu2005}, this change in luminance is likely due to temperature effects.
The absence of significant pressure fluctuations, coupled with reduced luminance in the PSP images, implies a higher temperature state. 
Therefore, the area observed at $x/c = 0.7$, $y/c = 0$ in Fig. \ref{fig:psp_1_p_1} can be interpreted as a high-temperature region.
It is known that focusing laser pulses creates a high-temperature, low-density area when laser plasma occurs \citep{Bright2018}, and it is believed that this phenomenon was captured in the current PSP measurements.

The advection speed of the high-temperature, low-density area, calculated from the PSP-measured time-series pressure fluctuation fields, was nearly identical to the freestream velocity at 22.5 m/s.
This similarity in speed arises because the laser pulses were focused outside of the boundary layer on the wing surface in this experiment.
It is believed that when this high-temperature, low-density area contacts the wing surface, it influences the PSP's luminescence.
As observed in Fig. \ref{fig:psp_1_te}, pressure fluctuations are suppressed after the passage of this high-temperature, low-density area at each chord location, suggesting its impact on the flow field. 
However, the current PSP measurement data lacks sufficient detail to fully comprehend the changes in the TE noise flow structure, including the variations in the flow velocity distribution perpendicular to the wing surface and within the velocity boundary layer.
Further study is required to gain an understanding of this phenomenon.

\textbf{\subsection{Redevelopment process}
\label{subsec:redevelop}}
This section discusses the measurement and analysis of the redevelopment process of pressure fluctuations at the trailing edge, occurring between $t = 16.6$ and $26.6$ ms, after the suppression process.
Figure \ref{fig:psp_2_p} shows the visualization of pressure fluctuation fields at $t = 18$, $20$, and $26$ ms.
Similarly to Fig. \ref{fig:psp_1_p}, the regions between $x/c = 0.4$ to $0.6$ and $y/c = -0.3$ to $0.3$, which are masked in gray, were affected by the laser pulses. 
Particularly, the low pressure observed at $x/c = 0.5$, $y/c = 0$ in Figs. \ref{fig:psp_2_p_1} and \ref{fig:psp_2_p_2}, and the high pressure in Fig. \ref{fig:psp_2_p_3}, do not represent accurate pressure fluctuation values.
At $t = 18$ ms (Fig. \ref{fig:psp_2_p_1}), the pressure fluctuation structure remains suppressed in the area of $x/c = 0.85$ to $0.95$ and $y/c = -0.1$ to $0.1$.
By $t = 20$ ms (Fig. \ref{fig:psp_2_p_2}), the amplitude of pressure fluctuations in the previously suppressed area is redeveloping, with the structure of the pressure fluctuations beginning to reform from the trailing edge.
Finally, at $t = 26$ ms (Fig. \ref{fig:psp_2_p_3}), the structure of the pressure fluctuation has recovered to the same extent as the surrounding areas. 
Consequently, PSP measurements have successfully captured the redevelopment process of pressure fluctuations.

Focusing on the pressure fluctuation values measured with PSP, the comparison of pressure values obtained from two pressure transducers installed at the trailing edge and from the surrounding PSP, plotted over time, is shown for $x/c = 0.9$, $y/c = -0.2$ in Fig. \ref{fig:psp_2_p_k_1}, and for $x/c = 0.9$, $y/c = 0$ in Fig. \ref{fig:psp_2_p_k_2}.
During the redevelopment process, pressure fluctuations are well captured at both locations.
In particular, in Fig. \ref{fig:psp_2_p_k_2}, the process of increasing the amplitude of pressure fluctuations as they redevelop is effectively tracked.
Additionally, despite not being immediately after the laser pulse application, the missing values around $t = 16.6$ ms are due to the characteristics of the first image taken in a sequence. 
Specifically, this was because the dark level of the image was different from that of the other frames.

Next, we compare the significant measurement errors observed in Fig. \ref{fig:psp_2_p_k_2} during $t = 16.6$ to $18.0$ ms. 
At this period, the pressure transducer data indicate fluctuations of about 10 Pa, challenging for PSP measurement. 
A potential cause of this error could be differences in the characteristics of the captured frames.
As described in Section \ref{sec:data_pro}, in the time-series measurements with the high-speed camera, the initial frames show a change in dark level by a few luminance levels, necessitating dark subtraction at each frame timing to reduce this effect.
However, the variation in luminance values in the dark images, despite being captured with the same exposure time by the high-speed camera, suggests a change in image sensitivity. 
Since the first 1 to 20 frames in Fig. \ref{fig:dark_plot} correspond to the period $t = 16.6$ to $17.6$ ms in Fig. \ref{fig:psp_2_p_k_2}, it can be attributed to the influence of the captured frames.
Moreover, this effect becomes more noticeable at times when the pressure fluctuations are very small. 
Measurement of small pressure fluctuations over time requires an understanding of the characteristics of the high-speed camera used.

\begin{figure}
    \includegraphics{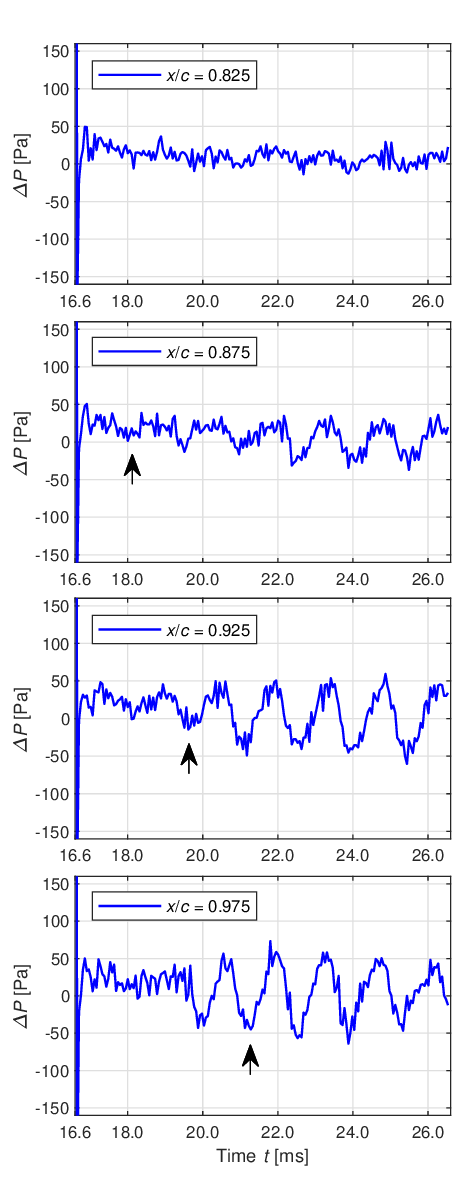}
    \caption{Time-series pressure fluctuations at various chord positions on the wing trailing edge ($x/c = 0.825$ to $0.975, y/c = 0$) during the redevelopment process}
    \label{fig:psp_2_te}
\end{figure}

For a detailed analysis of the redevelopment process of pressure fluctuations at the trailing edge, Fig. \ref{fig:psp_2_te} presents time-series pressure fluctuations at each chord location near the trailing edge ($x/c = 0.825$ to $0.975$), at $y/c = 0$, similar to Fig. \ref{fig:psp_1_te} during the suppression process.
From these results, it is observed that fluctuations begin around $t = 18$ ms at $x/c = 0.875$ and increase in amplitude as they move towards $x/c = 0.975$, as indicated by the black arrows in Fig. \ref{fig:psp_2_te}. 
In the time domain shown in Fig. \ref{fig:psp_2_te}, the amplitude of fluctuations at both $x/c = 0.875$ and $x/c = 0.925$ increases, but at $x/c = 0.975$, it starts to decrease after $t = 22$ ms.
When comparing the amplitude of pressure fluctuations at $x/c = 0.975$ during the redevelopment process (Fig. \ref{fig:psp_2_te}) with that during the suppression process (Fig. \ref{fig:psp_1_te}), it is observed that the amplitude before suppression is smaller than in the latter half of the redevelopment process.
It is anticipated that changes in the magnitude of pressure fluctuations will be observed at each chord location after $t = 26.6$ ms. 
Time-series data from the pressure transducer at $x/c = 0.9, y/c = 0$ (Fig. \ref{fig:timing}) further confirm that, during the measurement period of the redevelopment process, pressure fluctuations increase, reaching levels similar to those observed before suppression around $t = 29$ ms.

PSP measurements near the trailing edge within this measurement period revealed a transitional process not yet in a periodic steady state. 
Detailed pressure fluctuations at each chord location suggest that during the redevelopment process of TE noise, fluctuations develop from upstream but increase in magnitude from downstream to upstream. 

These measurement results demonstrate that time-resolved phase-lock PSP measurements can provide data to discuss the generation mechanism of TE noise pressure fluctuations.

\ \\

\section{Conclusions}
\label{sec:conclusion}
In this study, we conducted time-resolved phase-lock pressure-sensitive paint measurements on the surface of an airfoil, specifically focusing on TE noise. 
Our primary objective was to evaluate the feasibility of employing laser pulses directed at the airfoil surface to mitigate TE noise. 
The PSP measurement enabled us to observe the evolution of small pressure fluctuation structures at the trailing edge of the airfoil. 
By employing a semiconductor pressure transducer as a reference point, we captured multiple PSP images at a fixed frame rate of the high-speed camera, synchronized to the phase of pressure fluctuations.
Repetition of this process in several sets facilitated the collection of time-series images for phase averaging.
The principal conclusions are summarized as follows.
\begin{enumerate}
\item  The ensemble averaging of acquired image data at each phase timing enabled the visualization of small pressure fluctuations in the time domain, with an amplitude of 50 Pa, which would have been undetectable by PSP's single-shot method. 
Approximately 900 ensemble averages were required, and careful consideration of the dark level in each frame timing was essential.
\item Utilizing an \textit{in situ} method based on ensemble-averaged data of PSP images, we quantitatively captured the suppression and redevelopment process of pressure fluctuations ($\pm$ 50 Pa) with an accuracy of about 15 Pa. 
\item The temperature dependency of PSP allowed us to observe the formation of a high-temperature, low-density region following the laser pulse application, moving downstream at the freestream velocity. 
Subsequently, pressure fluctuations were observed to be suppressed several milliseconds after this region passed over the trailing edge.
\item Furthermore, our high spatial resolution time-series measurements with PSP elucidated the redevelopment process of TE noise after suppression at the trailing edge. We observed disturbances redeveloping upstream of the wing's trailing edge, with the amplitude of pressure fluctuations transiently recovering from the trailing edge towards the upstream direction.
\end{enumerate}

\section*{Appendix}
\phantomsection
\label{sec:appendix}
\addcontentsline{toc}{section}{Appendix}
\textbf{\subsection*{Evaluation of excitation light source stability}}
The stability of the UV-LED excitation light source (IL-106X, Hardsoft) was assessed, utilizing the same model as used during the wind tunnel tests. 
A cold filter (SC0451, Asahi Spectra, cut-off wavelength: 450 nm) was mounted on the UV-LED, and UV-LED was illuminated at the same power level as used during the tests.
Emissions from the UV-LED were recorded with a photodetector (C6386-01, Hamamatsu Photonics) and a digital oscilloscope (GR-7000, Keyence) at a sampling rate of 200 kHz over a 250 s period. 
Time-series data recorded during illumination were processed by subtracting the average value obtained when there was no illumination.

\begin{figure}[ht]
    \includegraphics{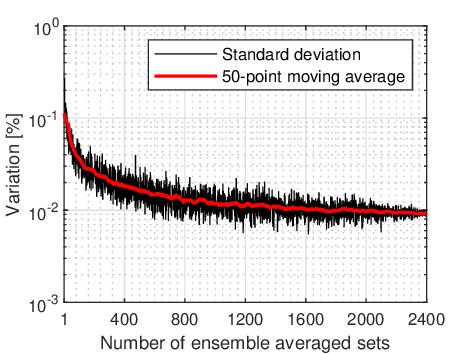}
    \caption{The effect of the number of ensemble averages on the variation in the excitation light source}
    \label{fig:led_var}
\end{figure}

From the processed time-series data, the standard deviation relative to the mean was calculated, revealing a variation of 0.4\%. 
Time-series data from the UV-LED during 10 ms measurement intervals—aligned with the wind tunnel test sequence (measuring once every 70 cycles of the 679 Hz TE noise frequency)—were extracted.
Approximately 2400 datasets were prepared and subjected to ensemble averaging, with datasets being randomly selected for averaging. 
A graph depicting the relative variation percentage as a function of the number of ensemble averages is presented in Fig. \ref{fig:led_var}. 
The black line represents the calculated variation percentages obtained through random sampling, while the red line shows these data smoothed by a 50-point moving average. 
It was observed that an increase in the number of ensemble averages further minimized the impact of variations from the UV-LED light source. 
After 1150 ensemble averages, the variation was found to be approximately 0.01\%.

\begin{figure}[ht]
    \includegraphics{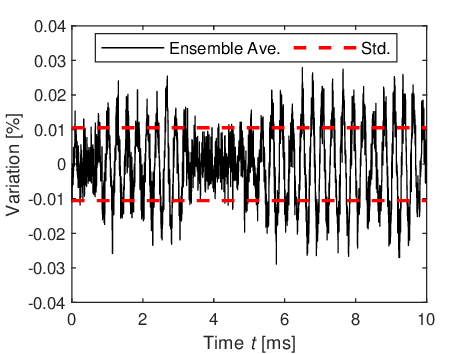}
    \caption{Time-series variation waveform of the excitation light source intensity after 1150 ensemble averages}
    \label{fig:led_ensemble}
\end{figure}

\begin{figure}[ht]
    \includegraphics{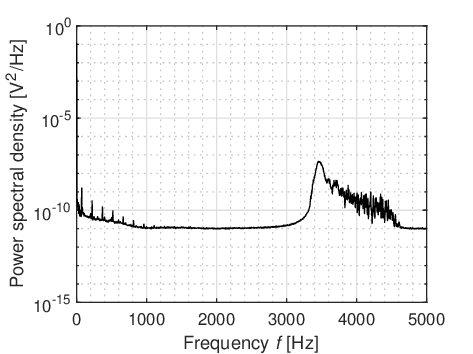}
    \caption{Power spectral density of the excitation light source time-series intensity}
    \label{fig:led_psd}
\end{figure}

Subsequent analyses focused on the output waveform over a 10 ms interval following ensemble averaging. 
Figure \ref{fig:led_ensemble} illustrates the time-series variation percentages over a 10 ms interval after 1150 ensemble averages, with the standard deviation indicated by a red dotted line. 
The observed variation within the 10 ms time series was 0.01\%, indicating minimal fluctuation in the UV-LED. 
The waveform contained high-frequency oscillations at approximately 3.5 kHz. 
Figure \ref{fig:led_psd} presents the power spectral density calculated from the measured time series data, showing multiple peaks in the low-frequency range (below 1.2 kHz) and a broad peak in the high-frequency range (3.5 to 4.5 kHz). 
Although ensemble averaging has reduced the influence of the low-frequency components, the broad high-frequency components have not been fully attenuated. 
These findings demonstrate that ensemble averaging effectively reduces noise in the excitation light source within this measurement sequence.

\section*{Declarations}
{\small \textbf{Ethics approval and consent to participate}
Not applicable}
\\
\\
{\small \textbf{Consent for publication}
Not applicable}
\\
\\
{\small \textbf{Availability of data and materials}
Not applicable}
\\
\\
{\small \textbf{Conflict of interest}
The authors have no conflict of interest to disclose.}
\\
\\
{\small \textbf{Funding}
The present study was supported by Japan Society for the Promotion of Science Grant-in-Aid for Scientific Research (KAKENHI) Grant Number 20K21043 and 22H01396.}
\\
\\
{\small \textbf{Author Contributions}
All authors were involved in the conceptualization of this study. 
MI, KK, and KN collaborated on developing the methodology.
KN also contributed resources.
MI was responsible for conducting the experiment, performing formal analysis and investigation, and preparing the original draft of the manuscript. 
MK was in charge of manuscript review, editing, funding acquisition, and supervision.
KK and KO also contributed to conducting the experiment.}
\\
\\
{\small \textbf{Acknowledgements}
Masato Imai gratefully acknowledges the support from the Support for Pioneering Research Initiated by the Next Generation of FLOuRISH Institute, Tokyo University of Agriculture and Technology (TUAT) granted by the Ministry of Education, Culture, Sports, Science and Technology (MEXT), Japan.
The authors would like to thank Mr. Tsutomu Nakajima of the Japan Aerospace Exploration Agency, for preparing the wing model with PSP coating.
The authors also thank Mr. Koichi Suzuki, Mr. Hideki Yamaya, and Mr. Tomoyuki Honma of the IHI Aerospace Engineering, for their technical support regarding the operation of the laser.
}
\\

\bibliographystyle{spbasic}      
\bibliography{ref_all}   

\begin{thebibliography}{26}
\providecommand{\natexlab}[1]{#1}
\providecommand{\url}[1]{{#1}}
\providecommand{\urlprefix}{URL }
\expandafter\ifx\csname urlstyle\endcsname\relax
  \providecommand{\doi}[1]{DOI~\discretionary{}{}{}#1}\else
  \providecommand{\doi}{DOI~\discretionary{}{}{}\begingroup \urlstyle{rm}\Url}\fi
\providecommand{\eprint}[2][]{\url{#2}}

\bibitem[{Asai and Yorita(2011)}]{Asai2011}
Asai K, Yorita D (2011) Unsteady {PSP} measurement in low-speed flow - overview of recent advancement at {T}ohoku {U}niversity. In: 49th AIAA Aerospace Sciences Meeting including the New Horizons Forum and Aerospace Exposition, \doi{10.2514/6.2011-847}

\bibitem[{Bertsch et~al(2019)Bertsch, Snellen, Enghardt, and Hillenherms}]{Bertsch2019}
Bertsch L, Snellen M, Enghardt L, Hillenherms C (2019) Aircraft noise generation and assessment: executive summary. CEAS Aeronautical Journal 10:3--9, \doi{10.1007/s13272-019-00384-3}

\bibitem[{Bright et~al(2018)Bright, Tichenor, Kremeyer, and Wlezien}]{Bright2018}
Bright A, Tichenor N, Kremeyer K, Wlezien R (2018) Boundary-layer separation control using laser-induced air breakdown. AIAA Journal 56:1472--1482, \doi{10.2514/1.J055272}

\bibitem[{Crafton et~al(2017)Crafton, Stanfield, Rogoshchenkov, and Schmit}]{Crafton2017}
Crafton J, Stanfield S, Rogoshchenkov N, Schmit R (2017) Investigation of passive flow control of cavity acoustics using dynamic pressure-sensitive paint. In: AIAA SciTech Forum - 55th AIAA Aerospace Sciences Meeting, \doi{10.2514/6.2017-1178}

\bibitem[{Desquesnes et~al(2007)Desquesnes, Terracol, and Sagaut}]{Desquesnes2007}
Desquesnes G, Terracol M, Sagaut P (2007) Numerical investigation of the tone noise mechanism over laminar airfoils. Journal of Fluid Mechanics 591:155--182, \doi{10.1017/S0022112007007896}

\bibitem[{Elliott et~al(2022)Elliott, Hasegawa, Sakaue, and Leonov}]{Elliott2022}
Elliott S, Hasegawa M, Sakaue H, Leonov S (2022) Shock-dominated flow control by plasma array: Pressure analysis including pressure-sensitive paint visualization. Experimental Thermal and Fluid Science 131, \doi{10.1016/j.expthermflusci.2021.110522}

\bibitem[{Gardner et~al(2014)Gardner, Klein, Sachs, Henne, Mai, and Richter}]{Gardner2014}
Gardner AD, Klein C, Sachs WE, Henne U, Mai H, Richter K (2014) Investigation of three-dimensional dynamic stall on an airfoil using fast-response pressure-sensitive paint. Experiments in Fluids 55:1--14, \doi{10.1007/s00348-014-1807-4}

\bibitem[{Gregory et~al(2006)Gregory, Sullivan, Wanis, and Komerath}]{Gregory2006}
Gregory JW, Sullivan JP, Wanis SS, Komerath NM (2006) Pressure-sensitive paint as a distributed optical microphone array. The Journal of the Acoustical Society of America 119:251--261, \doi{10.1121/1.2140935}

\bibitem[{Gregory et~al(2013)Gregory, Sakaue, Liu, and Sullivan}]{Gregory2013}
Gregory JW, Sakaue H, Liu T, Sullivan JP (2013) Fast pressure-sensitive paint for flow and acoustic diagnostics. Annual Review of Fluid Mechanics 46:303--330, \doi{10.1146/annurev-fluid-010313-141304}

\bibitem[{Gößling et~al(2020)Gößling, Ahlefeldt, and Hilfer}]{Goessling2020}
Gößling J, Ahlefeldt T, Hilfer M (2020) Experimental validation of unsteady pressure-sensitive paint for acoustic applications. Experimental Thermal and Fluid Science 112, \doi{10.1016/j.expthermflusci.2019.109915}

\bibitem[{Inasawa et~al(2013)Inasawa, Ninomiya, and Asai}]{Inasawa2013}
Inasawa A, Ninomiya C, Asai M (2013) Suppression of tonal trailing-edge noise from an airfoil using a plasma actuator. AIAA Journal 51:1695--1702, \doi{10.2514/1.J052203}

\bibitem[{Liu and Sullivan(2005)}]{Liu2005}
Liu T, Sullivan JP (2005) Pressure and {T}emperature {S}ensitive {P}aints. Springer, \doi{10.1007/b137841}

\bibitem[{Liu et~al(2001)Liu, Guille, and Sullivan}]{Liu2001}
Liu T, Guille M, Sullivan JP (2001) Accuracy of pressure-sensitive paint. AIAA Journal 39:103--112, \doi{10.2514/2.1276}

\bibitem[{Molin(2019)}]{Molin2019}
Molin N (2019) Airframe noise modeling and prediction. CEAS Aeronautical Journal 10:11--29, \doi{10.1007/s13272-019-00375-4}

\bibitem[{Nakakita(2011)}]{Nakakita2011}
Nakakita K (2011) Unsteady pressure measurement on naca0012 model using global low-speed unsteady psp technique. In: 41st AIAA Fluid Dynamics Conference and Exhibit, \doi{10.2514/6.2011-3901}

\bibitem[{Nakakita(2013)}]{Nakakita2013}
Nakakita K (2013) Detection of phase and coherence of unsteady pressure field using unsteady psp measurement. In: AIAA Ground Testing Conference, \doi{10.2514/6.2013-3124}

\bibitem[{Nash et~al(1999)Nash, Lowson, and McAlpine}]{Nash1999}
Nash EC, Lowson MV, McAlpine A (1999) Boundary-layer instability noise on aerofoils. Journal of Fluid Mechanics 382:27--61, \doi{10.1017/S002211209800367X}

\bibitem[{Noda et~al(2018)Noda, Nakakita, Wakahara, and Kameda}]{Noda2018}
Noda T, Nakakita K, Wakahara M, Kameda M (2018) Detection of small-amplitude periodic surface pressure fluctuation by pressure-sensitive paint measurements using frequency-domain methods. Experiments in Fluids 59, \doi{10.1007/s00348-018-2550-z}

\bibitem[{Ogura et~al(2023)Ogura, Kojima, Imai, Konishi, Nakakita, and Kameda}]{Ogura2023}
Ogura K, Kojima Y, Imai M, Konishi K, Nakakita K, Kameda M (2023) Reduction in airfoil trailing-edge noise using a pulsed laser as an actuator. Actuators 12, \doi{10.3390/act12010045}

\bibitem[{Paterson et~al(1973)Paterson, Vogt, Fink, and Munch}]{Paterson1973}
Paterson RW, Vogt PG, Fink MR, Munch CL (1973) Vortex noise of isolated airfoils. Journal of Aircraft 10:296--302, \doi{10.2514/3.60229}

\bibitem[{Peng and Liu(2020)}]{Peng2020}
Peng D, Liu Y (2020) Fast pressure-sensitive paint for understanding complex flows: from regular to harsh environments. Experiments in Fluids 61, \doi{10.1007/s00348-019-2839-6}

\bibitem[{Simon et~al(2016)Simon, Fabbiane, Nemitz, Bagheri, Henningson, and Grundmann}]{Simon2016}
Simon B, Fabbiane N, Nemitz T, Bagheri S, Henningson DS, Grundmann S (2016) In-flight active wave cancelation with delayed-x-lms control algorithm in a laminar boundary layer. Experiments in Fluids 57, \doi{10.1007/s00348-016-2242-5}

\bibitem[{Sugioka et~al(2018)Sugioka, Numata, Asai, Koike, Nakakita, and Nakajima}]{Sugioka2018}
Sugioka Y, Numata D, Asai K, Koike S, Nakakita K, Nakajima T (2018) Polymer/ceramic pressure-sensitive paint with reduced roughness for unsteady measurement in transonic flow. AIAA Journal 56:2145--2156, \doi{10.2514/1.j056304}

\bibitem[{Wylie et~al(2021)Wylie, Mishra, and Amitay}]{Wylie2021}
Wylie JD, Mishra S, Amitay M (2021) Tollmien–schlichting wave control on an airfoil using dynamic surface modification. AIAA Journal 59:2890--2900, \doi{10.2514/1.J060233}

\bibitem[{Yamamoto et~al(2019)Yamamoto, Hayama, Kumada, and Hayashi}]{Yamamoto2019}
Yamamoto K, Hayama K, Kumada T, Hayashi K (2019) A flight demonstration for airframe noise reduction technology. CEAS Aeronautical Journal 10:77--92, \doi{10.1007/s13272-019-00376-3}

\bibitem[{Yorita et~al(2010)Yorita, Nagai, Asai, and Narumi}]{Yorita2010}
Yorita D, Nagai H, Asai K, Narumi T (2010) Unsteady psp technique for measuring naturally-disturbed periodic phenomena. In: 48th AIAA Aerospace Sciences Meeting Including the New Horizons Forum and Aerospace Exposition, \doi{10.2514/6.2010-307}

\end{thebibliography}


\begin{thebibliography}{}
%
%
\bibitem{RefJ}
Author, Article title, Journal, Volume, page numbers (year)
\bibitem{RefB}
Author, Book title, page numbers. Publisher, place (year)
\end{thebibliography}

\end{document}